\documentclass[english]{elsarticle}

\usepackage[english]{babel}
\usepackage{amssymb,amsmath,amsthm,color}

\newcommand{\R}{\mathbb{R}}
\newcommand{\Z}{\mathbb{Z}}

\newcommand{\e}{\mathrm{e}}

\journal{Journal of \LaTeX\ Templates}

\begin{document}

\begin{frontmatter}

\title{Quantum graphs with vertices of a preferred orientation}

\author[mymainaddress,mysecondaryaddress]{Pavel Exner\corref{mycorrespondingauthor}}
\ead{exner@ujf.cas.cz}
\author[mymainaddress]{Milo\v{s} Tater}
\ead{tater@ujf.cas.cz}

\cortext[mycorrespondingauthor]{Corresponding author}
\address[mymainaddress]{Department of Theoretical Physics, Nuclear Physics Institute, Czech Academy of Sciences, 25068 \v Re\v z near Prague, Czechia}
\address[mysecondaryaddress]{Doppler Institute for Mathematical Physics and Applied Mathematics, Czech Technical University, B\v rehov\'a 7, 11519 Prague, Czechia}

\begin{abstract}
Motivated by a recent application of quantum graphs to model the anomalous Hall effect we discuss quantum graphs the vertices of which exhibit a preferred orientation. We describe an example of such a vertex coupling and analyze the corresponding band spectra of lattices with square and hexagonal elementary cells showing that they depend heavily on the network topology, in particular, on the degrees of the vertices involved.
\end{abstract}

\begin{keyword}
\texttt{Quantum graph, vertex coupling, preferred orientation, square lattice, hexagonal lattice, band spectrum. \MSC[2010] 81Q35, 34L40, 35J10}
\end{keyword}

\end{frontmatter}

\section{Introduction}

Quantum graphs represent an exceptionally fruitful concept both from the theoretical point of view as well as a tool for numerous applications -- for a review and a rich bibliography we refer to the monograph \cite{BK13}. The present letter is motivated by a recent application of the quantum graph technique to the anomalous Hall effect \cite{SK15}. The idea of this work is to model the motion of electrons in atomic orbitals by a network of rings with a $\delta$ coupling in their junctions; the one considered in \cite{SK15} is topologically equivalent to a square lattice giving rise to the Kronig-Penney-type spectrum \cite{Ex95}.

The model is simple and elegant but it has a drawback. In the real situation only atomic orbitals with particular angular momentum values are involved; to model such a situation in the quantum-graph setting one has to break the time-reversal invariance by assuming that electrons move on the rings in one direction only. Since this cannot be justified from the first principles, one is inspired to think how quantum graphs with a preferred orientation may look like. It is clear that restriction cannot be imposed on the edges on which the particle moves as on one-dimensional line segments. On the other hand, the vertex coupling offers such a possibility. While the couplings used typically in various models, the Kirchhoff one and more generally the $\delta$ coupling, as well as various versions of the $\delta'$ coupling and others, are time-reversal invariant, the general self-adjoint conditions given by \eqref{kschrader} below lose this property if the matrices $A,B$ are nonreal. The question is whether some couplings in this broad class can be associated with a rotational motion in lattice models of the mentioned type. Our goal is to introduce and analyze the simplest example of that type. We focus here on its properties rather to an application to particular physical effects; this would require to employ a more general class of such couplings with parameters that will allow, in particular, to tune the `rotation'.

The coupling we are going to discuss will be introduced in the next section where we will derive the appropriate spectral and scattering properties of a star-graph system. Then, in Sections~\ref{s:square} and \ref{s:hexagon}, we will analyze spectra of periodic lattices with the square and hexagon basic cells, respectively. Our main observation is that the transport properties of these systems depend heavily on the lattice topology, in particular, on the vertex degree of its junctions.

\section{Vertex coupling with a preferred orientation} 

Consider a star graph with $N$ semi-infinite edges which meet at a single vertex. The state Hilbert space associated with it is $\bigoplus_{j=1}^N L^2(\R_+)$, the elements of which are $\Psi=\{\psi_j\}$, and the Hamiltonian of the system in the absence of external fields is negative Laplacian, $H\{\psi_j\} = \{-\psi_j''\}$, where as usual we employ the units in which $\hbar=2m=1$. To make $H$ self-adjoint, one has to specify its domain by imposing suitable matching conditions to the boundary values of the functions $\{\psi_j\} \in \bigoplus_{j=1}^N H^2(\R_+)$. It is well known \cite{KoS99} that in general such conditions can be written as
\begin{equation} \label{kschrader}
A\Psi(0+)+B\Psi'(0+)=0\,,
\end{equation}
where the $N\times N$ matrices $A,B$ are such that the $N\times 2N$
matrix $(A|B)$ has the full rank and $A^*B$ is Hermitean;
alternatively one writes $A=U-I,\, B=i(U+I)$ where $U$ is an
$N\times N$ unitary matrix \cite{GG91, Ha00}.

Consider the reversion operator, $R\{\psi_j\} =\{\psi_{N+1-j}\}$. In contrast to the commonly used vertex couplings we are interested in those giving rise to Hamiltonians that \emph{do not} commute with $R$. In this paper we are going to consider the simplest example of this type in which the coupling exhibits `maximum rotation' at a fixed energy, here conventionally set to occur at the momentum $k=1\,$ (this fixes the momentum scale, of course, to change it one can put $k=p/p_0$ for a suitable $p_0$.). This is achieved by choosing
$$
U= \left( \begin{array}{ccccccc}
0 & 1 & 0 & 0 & \cdots & 0 & 0 \\ 0 & 0 & 1 & 0 & \cdots & 0 & 0 \\ 0 & 0 & 0 & 1 & \cdots & 0 & 0 \\ \cdots & \cdots & \cdots & \cdots & \cdots & \cdots & \cdots \\ 0 & 0 & 0 & 0 & \cdots & 0 & 1 \\ 1 & 0 & 0 & 0 & \cdots & 0 & 0
\end{array} \right)
$$
which is obviously unitary. In the component form, writing for simplicity $\psi_j=\psi_j(0+)$ and $\psi'=\psi'_j(0+),\: j=1,\cdots,N$, the matching conditions are
\begin{equation} \label{vertex}
 (\psi_{k+1}-\psi_k) + i(\psi'_{k+1}+\psi'_k) = 0\,, \quad k\in\Z\; (\mathrm{mod}\,N)\,;
\end{equation}
note that they are non-trivial only for $N\ge 3$. The non-invariance under $R$ is obvious, the first bracket changes sign under the reversion.

Let us look how the spectrum of the operator $H$ determined by the conditions \eqref{vertex} looks like. It not difficult to check that its essential component is absolutely continuous and coincides with the positive half of the real axis. As for discrete spectrum in the negative part, it follows from general principles that its dimension cannot exceed $N$. In fact the eigenvalues are easy to find: if we write the supposed solution as $\psi_i(x) = c_i \mathrm{e}^{-\kappa x}$ and plug this Ansatz into \eqref{vertex} we get a system of equations for the coefficients $c_i$. The requirement of its solvability yields the spectral condition
\[ 
(\kappa-i)^N + (-1)^{N-1} (\kappa+i)^N = 0\,,
\] 
which has solutions for any $N\ge 3$. Specifically, the star graph Hamiltonian $H$ has eigenvalues $-\kappa^2$, where
\begin{equation} \label{bs_ev}
\kappa= \tan \frac{\pi m}{N}
\end{equation}
with $m$ running through $1,\cdots,[\frac{N}{2}]$ for $N$ odd and $1,\cdots,[\frac{N-1}{2}]$ for $N$ even. Thus the discrete spectrum is always nonempty, in particular, $H$ has a single negative eigenvalue for $N=3,4$ which is equal to $-1$ and $-3$, respectively.

The quantity of interest in the continuous spectrum is the scattering matrix. It is straightforward to show \cite{BK13} that the on-shell S-matrix at the momentum $k$ is
\begin{equation} \label{smatrix}
S(k) = \frac{k-1 +(k+1)U}{k+1 +(k-1)U}\,.
\end{equation}
In particle, in accord with the construction the rotation is maximal for $k=1$ because then $S=U$ and one sees that a wave arriving at the vertex on the $k$th edge is diverted to the $(k-1)$th one, cyclically. It might seem that relation~\eqref{smatrix} implies that the transport becomes trivial at small and high energies, since $\lim_{k\to 0} S(k)=-I$ and $\lim_{k\to\infty} S(k)=I$.

However, more caution is needed; the formal limits may lead to a false result if the  $+1$ or $-1$ are eigenvalues of the matrix $U$. A counterexample can be found in vertices with Kirchhoff coupling. In that case $U$ has only $\pm 1$ as its eigenvalues and it is well known that the corresponding on-shell S-matrix is independent of~$k$ and it is \emph{not} a multiple of the identity.

Let us thus look at the right-hand side of \eqref{smatrix} more closely. A straightforward, even if a bit tedious computation yields the explicit form of $S(k)$: denoting
$$
\eta := \frac{1-k}{1+k}
$$
we have
\begin{equation} \label{smatrix_element}
S_{ij}(k) = \frac{1-\eta^2}{1-\eta^N} \left\{ -\eta\, \frac{1-\eta^{N-2}}{1-\eta^2}\,\delta_{ij} +
(1-\delta_{ij})\, \eta^{(j-i-1)(\mathrm{mod}\,N)} \right\}\,.
\end{equation}
In the lowest vertex degree cases formula \eqref{smatrix_element} yields
\begin{equation} \label{smatrix3}
S(k)= \frac{1+\eta}{1+\eta+\eta^2} \left( \begin{array}{ccc}
-\frac{\eta}{1+\eta} & 1 & \eta \\ \eta & -\frac{\eta}{1+\eta} & 1 \\ 1 & \eta & -\frac{\eta}{1+\eta}
\end{array} \right)
\end{equation}
and
\begin{equation} \label{smatrix4}
S(k)= \frac{1}{1+\eta^2} \left( \begin{array}{cccc}
-\eta & 1 & \eta & \eta^2 \\ \eta^2 & -\eta & 1 & \eta \\ \eta & \eta^2 & -\eta & 1 \\ 1 & \eta & \eta^2 & -\eta
\end{array} \right)
\end{equation}
for $N=3,4$, respectively. We see that $\lim_{k\to\infty} S(k)=I$
holds for $N=3$ and more generally for all odd $N$, while for the
even ones the limit is not a multiple of identity. This is is
obviously related to the fact that in the latter case $U$ has both
$\pm 1$ as its eigenvalues, while for $N$ odd $-1$ is missing.

\section{Square lattices} \label{s:square}

As indicated, our main topic in this letter are properties of periodic lattice graphs with the described reversion non-invariant coupling \eqref{vertex} at the vertices. Our first example is the square lattice of the edge length $\ell>0$. The system is periodic, hence its energy bands are obtained through investigation on the elementary-cell component of the Hamiltonian at a fixed value of the quasimomentum \cite[Chap.~4]{BK13}. We choose edge coordinates increasing in the direction up and right and use the abbreviation $\omega_j = \e^{i\theta_j},\, j=1,2$, for the Bloch phase factors. To match the four wave functions at a fixed vertex, conventionally labeled as $x=0$, we write the Ansatz
\begin{align}
\psi_1(x) &= a_1\e^{ikx} + b_1\e^{-ikx}\,, \nonumber \\
\psi_2(x) &= a_2\e^{ikx} + b_2\e^{-ikx}\,, \nonumber  \\[-.7em] \label{Ansatz} \\[-.7em]
\psi_3(x) &= \omega_1 \left( a_1\e^{ik(x+\ell)} + b_1\e^{-ik(x+\ell)} \right)\,, \nonumber \\
\psi_4(x) &= \omega_2 \left( a_2\e^{ik(x+\ell)} + b_2\e^{-ik(x+\ell)} \right)\,. \nonumber
\end{align}
Plugging the corresponding boundary values $\psi_j(0),\,\psi'_j(0),\, j=1,2,3,4$, into the conditions \eqref{vertex} taking into account that the derivatives there are taken in the direction away from the vertex, we get a system of linear equations for the coefficients $a_j,\,b_j$ which is solvable provided the determinant
\begin{equation} \label{determinant}
D \equiv D(\eta,\omega_1,\omega_2) = \left|
\begin{array}{cccc}
-1 & -\eta & \eta & 1 \\[.2em]
\omega_1\xi^2 & \omega_1\bar\xi^2\eta & -1 & -\eta
\\[.2em]
-\omega_1\xi^2\eta & -\omega_1\bar\xi^2 & \omega_2\xi^2 & \omega_2\bar\xi^2\eta \\[.2em]
\eta & 1 & -\omega_2\xi^2\eta & -\omega_2\bar\xi^2
\end{array}
\right|
\end{equation}
vanishes. Using the original momentum variable $k$ instead of $\eta$, this can be evaluated giving the expression
\begin{equation} \label{determinant}
D = 16i\,\e^{i(\theta_1+\theta_2)}\, k\, \sin k\ell \big[ (k^2-1) (\cos\theta_1 + \cos\theta_2) + 2(k^2+1) \cos k\ell \big]\,.
\end{equation}
Consequently, the spectrum of the lattice Hamiltonian consists of two parts:

\medskip

\noindent (a) \emph{infinitely degenerate eigenvalues}
$$
\lambda_m = \frac{\pi m}{\ell}\,,\quad m=0,1,2,\cdots\,,
$$
with the `elementary' eigenfunctions supported on single-square loops of the lattice. Let us remark that in the case of more common vertex couplings such as $\delta$ coupling these flat bands are often referred to as `Dirichlet' eigenvalues, because the eigenfunctions are composed of sine arcs vanishing at the lattice nodes. Here the name is not fitting having in mind, in particular, that the spectrum includes the point $k=0$ where the elementary eigenfunctions are constant on each four-edge square loop.

\medskip

\noindent (b) \emph{absolutely continuous bands:} they are determined by the spectral condition implied by \eqref{determinant} which reads
$$
\cos k\ell = \frac12 (\cos\theta_1 + \cos\theta_2) \frac{1-k^2}{1+k^2}\,,
$$
or alternatively
\begin{equation} \label{speccon}
\cos k\ell = \frac{1-k^2}{1+k^2}\, \cos\frac{\theta_1+\theta_2}{2}\, \cos\frac{\theta_1-\theta_2}{2}\,.
\end{equation}
This is not all, however. Similarly to the star graph the lattice Hamiltonian is not positive, we have to consider also negative energies corresponding to $k=i\kappa$ with $\kappa>0$, or replacing the trigonometric functions in \eqref{Ansatz}. This yields the spectral condition
\begin{equation} \label{speccon<0}
\cosh \kappa\ell = \frac{1+\kappa^2}{1-\kappa^2}\, \cos\frac{\theta_1+\theta_2}{2}\, \cos\frac{\theta_1-\theta_2}{2}\,.
\end{equation}

Let us look more closely at properties of the band spectrum. We denote
$$
c_\theta:= \cos\frac{\theta_1+\theta_2}{2}\,
\cos\frac{\theta_1-\theta_2}{2}\,;
$$
this quantity ranges though $[-1,1]$ as the quasimomentum $\vartheta= \frac1\ell \theta$ runs through the Brillouin zone $\big(-\frac{\pi}{\ell}, \frac{\pi}{\ell}\big]\times\big(-\frac{\pi}{\ell},\frac{\pi}{\ell}\big]$. Let us start with the \emph{negative band}:

\begin{itemize}

\item negative spectrum is \emph{never empty}; note that it is determined by the intersection of the function $\kappa \mapsto \cosh \kappa\ell$ with the region bordered from below and above by the curves $\kappa \mapsto \pm \frac{1+\kappa^2}{1-\kappa^2}$. This means, in particular, that the energy $-1$ corresponding to $\kappa=1$ belongs to the spectrum for any $\ell$ and $\inf\sigma(H)<-1$,

\item for $\ell>2$ the band is strictly negative, i.e. its upper edge is negative,

\item for large $\ell$ the negative band is exponentially narrow being approximately $[-1-2\e^{-\ell},-1+2\e^{-\ell}]$, up to an $\mathcal{O}(\e^{-2\ell})$ error. Note that this an expected behavior because the negative band is related to the eigenvalue \eqref{bs_ev} and the transport in the negative part of the spectrum means tunneling between the vertices which becomes more difficult as the edges lengthen,

\item on the other hand, for $\ell\le 2$ the negative band extends to zero,

\item the spectral threshold decreases as $\ell$ decreases, the corresponding solution to \eqref{speccon<0} being $\kappa = \sqrt{\frac2\ell} + \mathcal{O}(1)$ giving $\inf\sigma(H) = -\frac2\ell + \mathcal{O}(\ell^{-1/2})$ as $\ell\to 0$.

\end{itemize}

\noindent In a similar way, the \emph{positive band spectrum} is by
\eqref{speccon} determined by the intersection of the function $k
\mapsto \cos k\ell$ with the region bordered from below and above by
the curves $k \mapsto \pm \frac{1-k^2}{1+k^2}$. We see that

\begin{itemize}

\item the number of open gaps is always infinite,

\item the gaps are centered around the points $\frac{\pi m}{\ell}\:$ marking the flat bands except the lowest one; note that even if $m=1,2,\cdots$ it is still not appropriate to use the name Dirichlet because the `elementary' eigenfunction here have zero \emph{derivatives} at the vertices -- it would be thus more appropriate to speak about \emph{Neumann eigenvalues},

\item for $\ell\ge 2$ the first positive band starts at zero, on the contrary, for $\ell<2$ the first positive band is separated from zero,

\item it is possible that one positive band degenerates to a point, i.e. an infinitely degenerate eigenvalue; this happens for
$$
\ell = \frac{\pi}{2}\left( m-\frac12 \right)\,,\quad m=1,2,\cdots\,,
$$

\item the gap width is asymptotically constant in the energy scale, similarly as in the case of a $\delta$ coupling: it is $\frac{4}{\pi m} +
\mathcal{O}(m^{-2})$ in the momentum variable, i.e. $\frac{8}{\ell} + \mathcal{O}(m^{-1})$ in energy as $m\to\infty$.

\end{itemize}

\begin{figure}[htbp]
     \centering
     \includegraphics[clip, trim=1.5cm 6.8cm 4cm 15cm,width=0.8\textwidth]{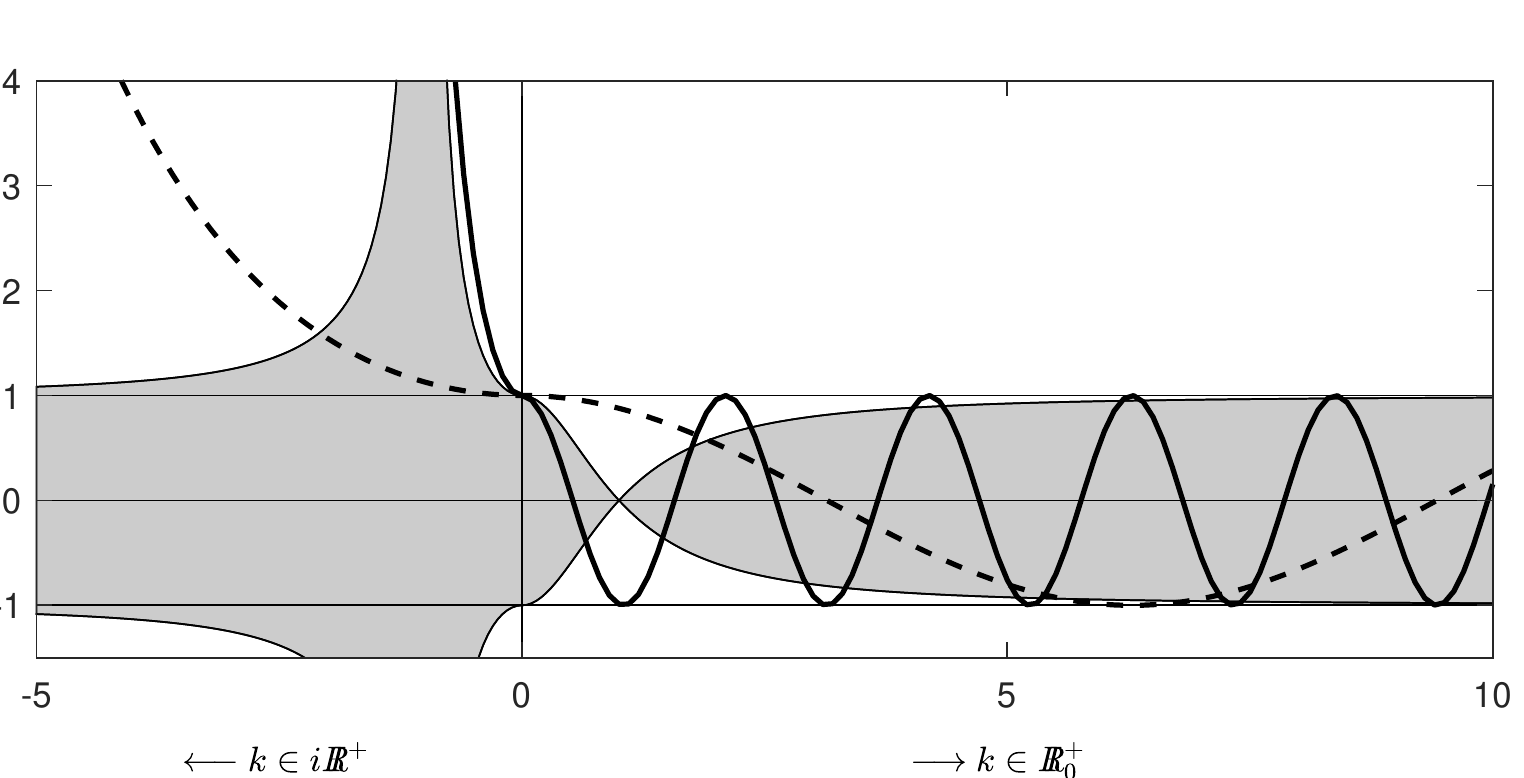}
     \caption{Spectral condition solution for the square lattice. The full line corresponds to $\ell=\frac32$, the dashed one to $\ell=\frac14$.}\label{soln_square}
\end{figure}

\noindent A graphical representation of the solution to spectral
conditions \eqref{speccon} and \eqref{speccon<0} is sketched in
Fig~\ref{soln_square} for two values of $\ell$; to put everything
into one picture, the right halfline shows the variable $k$, the
left one $\kappa$. The intersection describing the negative band
for $\ell=\frac32$ lays outside the picture area.

\section{Hexagonal lattices} \label{s:hexagon}

\begin{figure}[htbp]
     \centering
     \includegraphics[clip, trim=1.5cm 15.9cm 10.6cm 6.5cm,
width=0.45\textwidth]{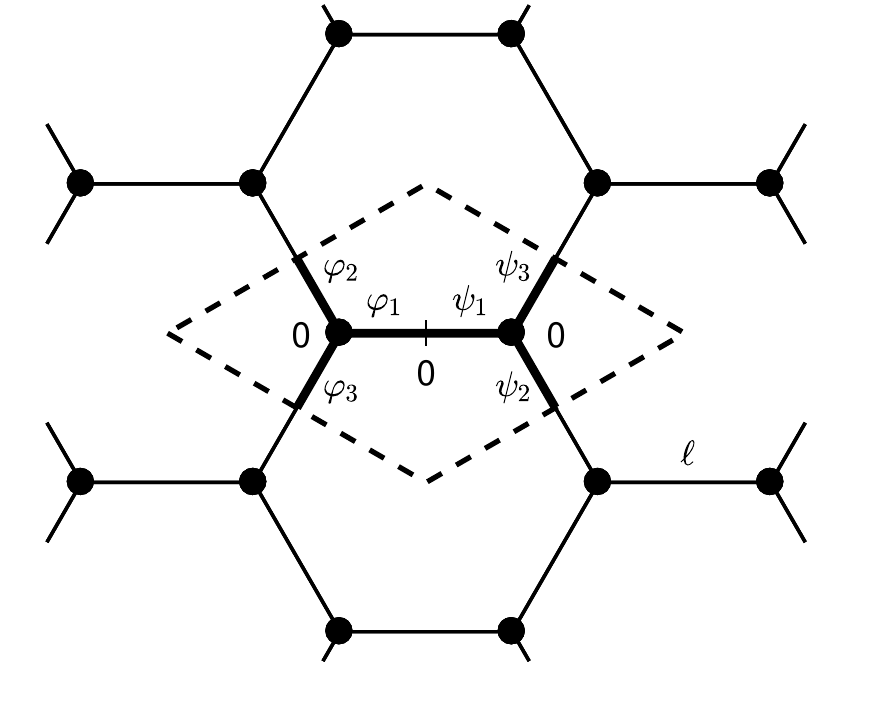}
     \caption{An elementary cell of the hexagon network}\label{elemcell}
\end{figure}

The method is in general the same but the Floquet-Bloch analysis becomes more complicated here because the elementary cell of a hexagonal lattice contains two vertices, cf. Fig.~\ref{elemcell}.
We choose the coordinates to increase `from left to right', and correspondingly, the Ansatz~\eqref{Ansatz} is now replaced by
\begin{align}
\psi_1(x)&=C_1^+\e^{\i k x}+C_1^-\e^{-\i k x} \nonumber \\
\psi_2(x)&=C_2^+\e^{\i k x}+C_2^-\e^{-\i k x} \label{psi} \\
\psi_3(x)&=C_3^+\e^{\i k x}+C_3^-\e^{-\i k x} \nonumber
\end{align}
for $x\in[0,\frac12\ell]$ and
\begin{align}
\varphi_1(x)&=D_1^+\e^{\i k x}+D_1^-\e^{-\i k x} \nonumber \\\
\varphi_2(x)&=D_2^+\e^{\i k x}+D_2^-\e^{-\i k x} \label{phi} \\
\varphi_3(x)&=D_3^+\e^{\i k x}+D_3^-\e^{-\i k x} \nonumber
\end{align}
for $x\in[-\frac12\ell,0]$. Naturally, the functions $\psi_1$ and $\varphi_1$ have to be matched smoothly, $\psi_1(0)=\varphi_1(0)$ and $\psi_1'(0)=\varphi_1'(0)$, which yields
\begin{equation}\label{CD1}
C_1^+=D_1^+\,,\quad C_1^-=D_1^-\,.
\end{equation}
Introducing $\xi =\e^{ik\ell/2}$ and $\omega_j = \e^{i\theta_j},\, j=1,2$, we get further
\begin{align}
D_2^+ =\xi^2\bar\omega_1 C_2^+\,,& \quad D_2^- =\bar\xi^{\,2}\bar\omega_1 C_2^-\,, \nonumber \\[-.6em] \label{CD23} \\[-.6em]
D_3^+ =\xi^2\bar\omega_2 C_3^+\,,& \quad D_3^- =\bar\xi^{\,2}\bar\omega_2 C_3^-\,, \nonumber
\end{align}
Next one has to match the function $\psi_j,\, j=1,2,3$, and $\varphi_j,\, j=1,2,3$, using the conditions \eqref{vertex} paying proper attention to signs coming from the directions in which the derivatives are taken. This yields
\begin{align*}
\psi_2(0) - \psi_1(\ell/2) + i\big(\psi'_2(0) - \psi'_1(\ell/2) \big) &= 0\,, \\
\psi_3(0) - \psi_2(0) + i\big(\psi'_3(0) + \psi'_2(0) \big) &= 0\,, \\
\psi_1(\ell/2) - \psi_3(0) + i\big(-\psi'_2(0) + \psi'_1(\ell/2) \big) &= 0\,, \\
\varphi_2(0) - \varphi_1(-\ell/2) + i\big(-\varphi'_2(0) + \varphi'_1(\ell/2) \big) &= 0\,, \\
\varphi_3(0) - \varphi_2(0) - i\big(\varphi'_3(0) + \varphi'_2(0) \big) &= 0\,, \\
\varphi_1(-\ell/2) - \varphi_3(0) + i\big(\varphi'_1(-\ell/2) - \varphi'_3(0) \big) &= 0\,.
\end{align*}
Substituting here from \eqref{psi}--\eqref{CD23} we get a system of six linear equations for the coefficients $C_j^\pm,\, j=1,2,3$. Computing the corresponding determinant we arrive at the spectral condition
$$
16i\,\e^{-i(\theta_1+\theta_2}\,k^2\sin k\ell\, \Big( 3 + 6k^2 - k^4 +4d_\theta(k^2-1) + (k^2+3)^2 \cos 2k\ell \Big) = 0\,,
$$
where
$$
d_\theta := \cos\theta_1 + \cos(\theta_1-\theta_2) + \cos\theta_2\,,
$$
which requires either $\sin k\ell=0$ or
\begin{equation} \label{hexcondition}
\cos 2k\ell = \frac{k^4 - 6k^2 - 3 - 4d_\theta(k^2-1)}{(k^2+3)^2}\,.
\end{equation}
As in the square-lattice case, this is accompanied with the negative-energy condition
\begin{equation} \label{hexcondition<0}
\cosh 2\kappa\ell = \frac{\kappa^4 + 6\kappa^2 - 3 + 4d_\theta(\kappa^2+1)}{(\kappa^2-3)^2}\,.
\end{equation}
The spectrum of the hexagonal-lattice Hamiltonian consists thus again of two parts:

\medskip

\noindent (a) \emph{infinitely degenerate eigenvalues}
$$
\lambda_m = \frac{\pi m}{\ell}\,,\quad m=0,1,2,\cdots\,,
$$
with the eigenfunctions composed of elementary ones supported by the loops of the lattice. One can speak of Neumann eigenvalues again, note that the hexagonal cell has an even number of edges, so the eigenfunction may keep switching sign at the vertices along the loop.

\medskip

\noindent (b) \emph{absolutely continuous bands} determined by the conditions \eqref{hexcondition} and \eqref{hexcondition<0}; we note that $d_\theta \in[-1,3]$ reaching the maximum in the center of the Brillouin zone and minimum at its edges. Let us start with the \emph{negative spectrum}:

\begin{itemize}

\item

it is again \emph{never empty} being determined by the intersection
of the function $\kappa \mapsto \cosh 2\kappa\ell$ with the region
bordered from below and above by the curves $\kappa \mapsto
\frac{g_\pm(\kappa)}{(\kappa^2-3)}$, where $g_+(\kappa) =
\kappa^4+18\kappa^2+9$ and $g_-(\kappa) = \kappa^4+2\kappa^2-7$.
This means, in particular, that $\inf\sigma(H)<-3$ and the negative
spectrum consists of \emph{two} bands below and above the energy
$-3$,

\item for $\ell>2/\sqrt{3}\approx 1.155$ the second band is strictly negative, i.e. its upper edge is negative; on the other hand, for $\ell\le 2/\sqrt{3}$ the negative band extends to zero,

\item for large $\ell$ the negative bands are exponentially narrow. The single vertex bound state energy \eqref{bs_ev} is again manifested; the bands are centered around it being of the size $\approx 8\,\mathrm{e}^{-\ell\sqrt{3}}$ being separated by a gap of the same size, all up to an $\mathcal{O}(\e^{-2\ell\sqrt{3}})$ error,

\item the first band decreases as $\ell$ decreases being $\big(-\frac{2}\ell, -\frac{2\sqrt{3}}\ell \big)$ up to an $\mathcal{O}(\ell^{-1/2})$ error as $\ell\to 0$.

\end{itemize}

\noindent Similarly the \emph{positive band spectrum} is determined
by the intersection of the function $k \mapsto \cos 2k\ell$ with the
region bordered from below and above by the curves $k \mapsto
\frac{h_\pm(k)}{(k^2+3)}$, where $h_+(k) = k^4-18k^2+9$ and $h_-(k)
= k^4-2k^2-7$. This implies that

\begin{itemize}

\item

the number of open gaps is always infinite. The first positive band
starts at zero if $\ell\le 2/\sqrt{3}$, otherwise a gap between it
and the second negative band opens,

\item at higher energies the \emph{bands} appear in pairs centered around the points $\frac{\pi m}{\ell}\:$ marking the flat bands (Neumann eigenvalues)

\item it is again possible that one positive band degenerates to a point, i.e. an infinitely degenerate eigenvalue; this time it happens for
$$
\ell = \left\{ \frac{\pi}{3}, \frac{2\pi}{3}\right\} \:
(\mathrm{mod} \,\pi)\,,
$$

\item at high energies gaps dominate the spectrum. The two bands around $k^2= \frac{\pi m}{\ell}$ has asymptotically the widths $\frac{4(\sqrt{3}-1)}{\ell} + \mathcal{O}(m^{-1})$ and the gap between them is $\frac{8}{\ell} + \mathcal{O}(m^{-1})$ as $m\to \infty$.

\end{itemize}

\begin{figure}[htbp]
     \centering
     \includegraphics[clip, trim=1.5cm 6.8cm 4cm 15cm,width=0.8\textwidth]{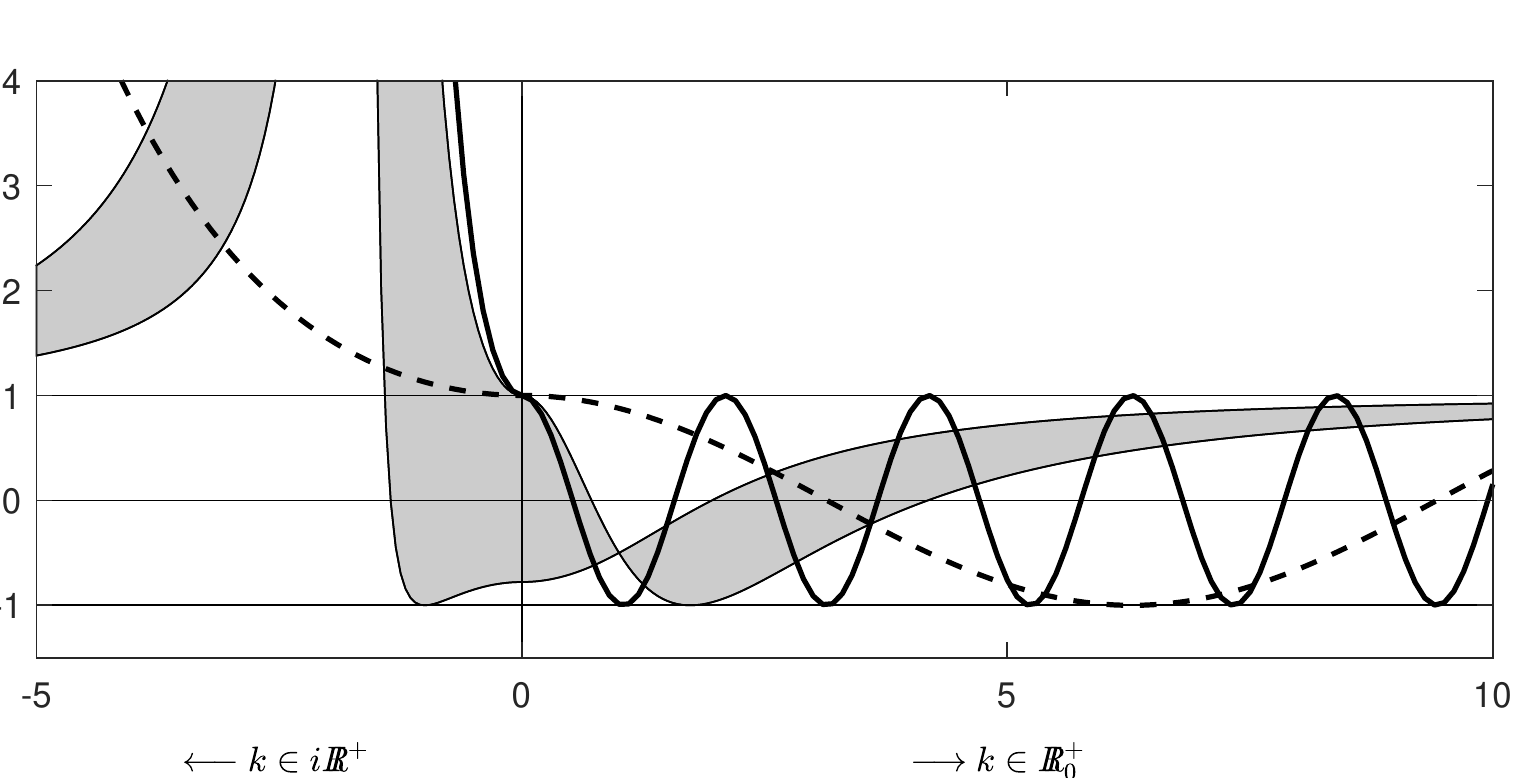}
     \caption{Spectral condition solution for the hexagonal lattice, the full and dashed lines respectively correspond to the same values as in Fig~\ref{soln_square}.}\label{soln_hexa}
\end{figure}

\noindent As in the previous case one can represent solution to
conditions \eqref{hexcondition} and \eqref{hexcondition<0}
graphically as sketched in Fig~\ref{soln_hexa} with same conventions
as above; the intersections marking the negative-energy solutions
for $\ell=\frac32$ lay again outside the picture area.

\section{Conclusions} 

It is instructive to compare the properties of the periodic lattices discussed in the previous two sections. Both exhibit flat bands, or infinitely degenerate eigenvalues. This effect, demonstrating one more time the invalidity of the unique continuation principle in quantum graphs, is well known, the specific feature here is that the corresponding eigenfunction components have Neumann rather than Dirichlet behavior. On the other hand, the two lattices sharply differ from the viewpoint of the absolutely continuous spectral component. The square one is `transport friendly', in the hexagon lattice bands occur in pairs and it is the gaps which dominate at high energies. It is obvious that these differences are related to the properties of the single vertex scattering matrices \eqref{smatrix3} and \eqref{smatrix4}: in the hexagon case the vertices are of degree three and the reflection dominates at high energies, while the degree-four vertices of the squre lattice distribute the particles evenly in the limit $k\to\infty$.

We note also that the coupling \eqref{vertex} was constructed to `maximize' the rotation. It would be useful to examine couplings that interpolate between \eqref{vertex} and some standard time-reversal invariant ones; this will be done in another paper.

\subsection*{Acknowledgments}
\noindent We thank Pavel St\v{r}eda for useful discussions. The research was supported by the Czech Science
Foundation (GA\v{C}R) within the project 17-01706S.


\end{document}